# Some Limitations of Dislocation Walls as Models for Plastic Boundary Layers

## Michael Zaiser* and Istvan Groma**

*Institute for Materials and Processes, School of Engineering, The University of Edinburgh, The King's Buildings, Sanderson Building, Edinburgh, EH9 3JL, UK
**Department of Materials Physics, Eötvös Loránd University, Pázmány Péter
s´étány 1/a, H-1117 Budapest, Hungary

**Abstract**. It has recently become popular to analyze the behavior of excess dislocations in plastic deformation under the assumption that such dislocations are arranged into walls with periodic dislocation spacing along the wall direction. This assumption is made plausible by the fact that periodic walls represent minimum energy arrangements for dislocations of the same sign, and it allows to use the analytically known short-ranged stress fields of such walls for analyzing the structure of plastic boundary layers. Here we show that unfortunately both the idea that dislocation walls are low-energy configurations and the properties of their interactions depend critically on the assumption of a periodic arrangement of dislocations within the walls. Once this assumption is replaced by a random arrangement, the properties of dislocation walls change completely.

## INTRODUCTION

Plastic boundary layers, i.e., transition regions between plastically deforming and non-deforming regions, are of importance for understanding a large number of phenomena in plasticity theory, such as the hardening of alloys by non-deformable inclusions, size effects occurring in plastic deformation of confined systems, and the deformation of polycrystals at low stresses where grain boundaries act as efficient dislocation obstacles. They are also of fundamental conceptual interest, since large strain gradients in the boundary region imply high dislocation densities, so that the interactions between individual dislocations become important even if one wants to describe the deformation process in a continuum setting. In this sense, the plastic boundary layer problem can be considered paradigmatic for the transition between discrete and continuum descriptions of dislocation plasticity [1].

Recently, it has become popular to model the structure of plastic boundary layers in terms of arrays of periodic dislocation walls, i.e., walls where dislocations of the same sign are arranged periodically above each other, similar to twist or tilt small-angle grain boundaries [2]. In physical terms, such an approach appears at first glance to be well motivated by the following arguments: (i) In the vicinity of impenetrable obstacles, the dislocation microstructure is dominated by 'geometrically necessary' dislocations of one sign piling up against the obstacle, which justifies to neglect 'statistically stored' dislocations of zero net Burgers vector; (ii) periodic walls are the lowest-energy configurations for dislocations of the same sign; (iii) for periodic walls, analytical expressions for the stress fields exist. These fields are of short range, which may justify an approximation of the interactions in terms of dislocation density gradients – an approximation which provides useful links with higher-order strain gradient plasticity; (iv) the presence of multiple walls (rather than a single wall which would be the absolute energy minimum for a given number of dislocations of the same sign) can be understood by the emission of dislocation processions from sources: The dislocations in such a procession are necessarily arranged one behind the other, and in absence of climb they cannot collapse into a single wall.

In this paper we are going to argue that the arguments (ii) and (iii) cease to hold if we avoid the unphysical assumption that the dislocations are emitted from an infinite array of periodically spaced sources. It is clear that this assumption can never be fulfilled in real materials unless we assume some Maxwell demon to put the dislocation sources on the requisite positions. If it is replaced by the more realistic assumptions of random dislocation source locations, and accordingly random slip plane spacings, then the properties of dislocation walls change dramatically. To show this, we first calculate the energy of a dislocation wall with random dislocation positions, and then we study the interactions between such walls.

## THE ENERGETICS OF RANDOM DISLOCATION WALLS

We consider a wall of infinite height running along the plane $x=0$. Edge dislocations of Burgers vector $\boldsymbol{b} = b\boldsymbol{e}_x$ are distributed randomly along this plane with average linear density $1/h$. This geometry corresponds to a plane-strain situation, hence the elastic energy density of an isotropic material can be written as

$$\mathcal{E} = \frac{1}{4G}\left[\left(\sigma_{xx} + \sigma_{yy}\right)^2(1-\nu) + 2\left(\sigma_{xy}^2 - \sigma_{xx}\sigma_{yy}\right)\right]. \tag{1}$$

Here, $G$ is the shear modulus and $\nu$ is Poisson's ratio. The total energy of the system is obtained by integrating Eq. (1) over the system volume:

$$E = \int_V \mathcal{E}d^2r = \frac{1-\nu}{4G}\int_V\left(\sigma_{xx} + \sigma_{yy}\right)^2 d^2r \ , \tag{2}$$

where we have used that, for an infinite system, the second and third terms on the right-hand side do not contribute to the total energy. This can be shown as follows: For plane-strain deformation, the stresses can be written as derivatives of the Airy stress function $\chi$: $\sigma_{xx} = \partial_y^2\chi, \sigma_{yy} = \partial_x^2\chi, \sigma_{xy} = -\partial_x\partial_y\chi$. The integral of the second term is

$$\frac{1}{2G}\int_V\left[\partial_x\partial_y\chi\partial_x\partial_y\chi - \partial_x^2\chi\partial_y^2\chi\right]d^2r \qquad . \tag{3}$$

Partially integrating the terms in the integral with respect to $x$ and $y$ shows that this integral contributes only surface terms to the total energy. These terms are negligible in the infinite-system limit.

The ensemble-averaged stress at any point is given by summing over the stress fields of the individual dislocations in the wall and averaging over the different realizations of the random dislocation positions:

$$\left\langle\sigma_{ij}(\boldsymbol{r})\right\rangle = \sum_n\left\langle\sigma_{ij}^{(n)}(\boldsymbol{r})\right\rangle \qquad , \tag{4}$$

where $\sigma_{ij}^{(n)}(\boldsymbol{r})$ is the $ij$ component of the stress created at $\boldsymbol{r}$ by the $n$th dislocation. The elastic energy density of the system depends on the averages of products $\sigma_{ij}(\boldsymbol{r})\sigma_{kl}(\boldsymbol{r})$ where $\{ij,kl\} \in [\{xx,xx\},\{xx,yy\},\{yy,yy\}]$. In evaluating these averages we use that the individual dislocation positions are independent random variables:

$$\left\langle\sigma_{ij}(\boldsymbol{r})\sigma_{kl}(\boldsymbol{r})\right\rangle = \left\langle\sum_n\sigma_{ij}^{(n)}(\boldsymbol{r})\sum_m\sigma_{kl}^{(m)}(\boldsymbol{r})\right\rangle = \left\langle\sum_n\sigma_{ij}^{(n)}(\boldsymbol{r})\sigma_{kl}^{(n)}(\boldsymbol{r})\right\rangle + \left\langle\sum_n\sigma_{ij}^{(n)}(\boldsymbol{r})\sum_{m\neq n}\sigma_{kl}^{(m)}(\boldsymbol{r})\right\rangle$$
$$= \left\langle\sum_n\sigma_{ij}^{(n)}(\boldsymbol{r})\sigma_{kl}^{(n)}(\boldsymbol{r})\right\rangle + \left\langle\sigma_{ij}(\boldsymbol{r})\right\rangle\left\langle\sigma_{kl}(\boldsymbol{r})\right\rangle - \sum_n\left\langle\sigma_{ij}^{(n)}(\boldsymbol{r})\right\rangle\left\langle\sigma_{kl}^{(n)}(\boldsymbol{r})\right\rangle. \tag{5}$$

We observe that the average stresses $<\sigma_{ij}(\boldsymbol{r})>$ and their products $<\sigma_{ij}(\boldsymbol{r})\sigma_{kl}(\boldsymbol{r})>$ depend on the $x$ coordinate only. The average single-dislocation stresses $<\sigma_{xx}^{(n)}(\boldsymbol{r})>$ and $<\sigma_{yy}^{(n)}(\boldsymbol{r})>$ become zero in the limit of infinite system size, since these stresses are antisymmetric functions of the $y$ coordinate. The same is true for the average total stresses $<\sigma_{xx}(\boldsymbol{r})>$ and $<\sigma_{yy}(\boldsymbol{r})>$ . With these observations only the first term in the second line of Eqn. (5) contributes to the system energy density, which thus reduces to

$$\mathcal{E} = \sum_n\left\langle\mathcal{E}^{(n)}\right\rangle. \tag{6}$$

Thus, the average energy density of the wall is nothing but the sum of the energy densities arising from the stress fields of the individual dislocations (i.e. the energy density of an array of non-interacting dislocations). This energy density is easily evaluated by inserting into Eq. (1) the stress field of a single-dislocation as given in [3] and averaging over the random $y$ positions of the dislocations.

We see from the above considerations that the infinite random wall does *not* represent a low-energy configuration. Its energy is (up to non-extensive terms that are negligible in the infinite system limit) exactly the same as the energy of the same number of isolated dislocations without screening interactions. Accordingly, the thermodynamic driving force towards forming such a wall, starting from a random dislocation arrangement, is zero.

So what about the common idea that dislocations of the same sign arrange into wall configurations to reduce their energy? An analysis of the correlations arising from the relaxation of random dislocation systems [4] under glide-only conditions indeed demonstrates that dislocations of the same sign preferentially arrange vertically above each other [5]. However, these screening correlations are of short range and become negligible above the scale of a few dislocation spacings. Thus, under glide conditions the relaxed configuration is not a single giant wall but a pattern of small correlated, wall-like clusters. If climb is possible, then the motion of dislocations perpendicular to the glide direction can lead to a regular arrangement providing a significant reduction in energy, thus providing an energetic incentive for the formation of extended walls. This is however not relevant for the analysis of plasticity boundary layers: Climb motion allows all geometrically necessary dislocations to form a single wall directly at the impenetrable boundary of the plastic domain, thus recovering the discontinuous solution of classical plasticity theory which, in the absence of climb, was kinematically unattainable for the dislocation system. Boundary layers, on the other hand, form precisely because of the kinematic constraints that in the absence of climb motion prevent dislocations to from reaching the boundary.

## THE INTERACTION OF RANDOM DISLOCATION WALLS

So what happens if dislocations emitted from randomly placed sources are pushed against an impenetrable boundary? How do they interact? Within the wall picture, we consider two random walls and evaluate their interaction under two different assumptions: (1) the positions of the dislocations in one wall are uncorrelated with respect to the positions of dislocations in the other wall; (2) the dislocations in both walls share common slip planes (e.g. because they have been emitted from the same sources). We consider a representative dislocation of the first wall. Without loss of generality we assume this dislocation to be located in the origin. The shear stress on the representative dislocation due to the dislocations in the second wall, which we assume to be located in the plane $x=d$, is given by

$$\sigma_{xy} = \sum_n \sigma_{xy}^{(n)}(\boldsymbol{r}^{(n)})$$ (7)

where the summation runs over the dislocation positions in the second wall. If the dislocation positions in both walls are mutually uncorrelated, the average stress experienced by the representative dislocation in the first wall is

$$\left\langle \sigma_{xy}(d) \right\rangle = \left\langle \sum_n \sigma_{xy}^{(n)}(d, y^{(n)}) \right\rangle = \left\langle \sum_n \int \delta(x-d)\delta(y-y^{(n)})\sigma_{xy}(x,y)d^2r \right\rangle$$

$$= \int \left\langle \sum_n \delta(y-y^{(n)}) \right\rangle \sigma_{xy}(d,y)d^2r = \frac{1}{h}\int \sigma_{xy}(d,y)dy = 0$$ (8)

Thus the interaction between two random walls with uncorrelated dislocation positions vanishes. The case where the representative dislocation shares a common slip planes with a 'correlated partner' in the other wall is constructed from the random case as follows: We remove a random dislocation (with label $k$) from the uncorrelated wall and place it at $y=0$. In this case, the average stress experienced by the representative dislocation is

$$\left\langle \sigma_{xy}(d) \right\rangle = \sigma_{xy}(d,0) + \left\langle \sum_{n \neq k} \sigma_{xy}^{(n)}(d, y^{(n)}) \right\rangle$$

$$= \sigma_{xy}(d,0) + \left\langle \sum_n \sigma_{xy}^{(n)}(d, y^{(n)}) \right\rangle - \left\langle \sigma_{xy}(d, y^{(k)}) \right\rangle = \sigma_{xy}(d,0)$$ (9)

since again the averages over the uncorrelated positions vanish. In other words, the total interaction force between two random walls with correlated positions is the same as the sum of the interaction forces of the isolated dislocations. Thus, the structure of a boundary layer composed of such walls is identical to that of an array of mutually non-interacting classical pile-ups.

## CONCLUSIONS

We have shown that most of the properties that make periodic dislocation walls attractive as 'building blocks' for the modelling of plasticity boundary layers (and more generally, as building blocks for continuum models of plasticity) depend crucially on the assumption of a periodic dislocation arrangement within the walls. If this assumption is replaced by an uncorrelated random arrangement (which is what we expect if the walls consist of dislocations emitted from randomly located sources) then the properties of dislocation walls change dramatically:

- o A periodic dislocation wall is a low-energy configuration where dislocation stress fields are screened on scales above the dislocation spacing in the wall. A random dislocation wall is energetically equivalent to the same number of isolated dislocations, i.e. the self-energy of dislocations in such a wall continues to diverge in proportion with the logarithm of crystal size.
- o Two periodic dislocation walls exhibit short-range repulsive interactions which exhibit a characteristic length scale proportional to the dislocation spacing in the wall. Two random dislocation walls either exhibit no interaction whatsoever (if the dislocations in both walls are not positionally correlated) or long-range interactions which are identical to those of isolated dislocations moving on the same slip plane (if each dislocation has a partner in the other wall that shares the same slip plane)

As there is no energetic incentive to form random walls, we cannot expect excess dislocations in a plasticity boundary layer to arrange into extended wall configurations. Even if they would do so, this would not change the long-range nature of their interactions. Walls can therefore not be considered as suitable building blocks for analysing the influence of short-range dislocation interactions in plasticity boundary layers – unless there would be a physical mechanism available to arrange the dislocations in a periodic manner.